\documentclass[12pt]{iopart}

\usepackage{graphicx}% Include figure files
\usepackage{bm}% bold math
\newcommand{\units}[1]{\:\mathrm{#1}}                                       % Einheitenformatierung: mit Abstand und nicht kursiv
\newcommand{\idx}[1]{_{\mathrm{#1}}}                                        % nicht-kursiv-stellen von Indizes in math. Ausdrücken
\begin{document}

\title[Sympathetic cooling of Yb and Rb]{Sympathetic cooling in a mixture of diamagnetic and paramagnetic atoms}

\author{S.~Tassy, N.~Nemitz, F.~Baumer, C.~H\"ohl, A.~Bat\"ar and A.~G\"orlitz}

\address{Institut f\"ur Experimentalphysik,
Heinrich-Heine-Universit\"at D\"usseldorf, Universit\"atsstra{\ss}e 1,
40225~D\"usseldorf, Germany}
\ead{axel.goerlitz@uni-duesseldorf.de}
\begin{abstract}
We have experimentally realized a hybrid trap for ultracold
paramagnetic rubidium and diamagnetic ytterbium atoms by combining
a bichromatic optical dipole trap for ytterbium with a
Ioffe-Pritchard-type magnetic trap for rubidium. In this hybrid
trap, we have investigated sympathetic cooling for five different ytterbium isotopes
through elastic collisions with rubidium. A strong dependence of the thermalization rate on the mass of the specific ytterbium isotope was observed.

\end{abstract}

\pacs{34.50.Cx, 37.10.De, 37.10.Gh}         % 37.10.Gh Atom traps and guides
                                % 34.50.Cx Elastic ultracold collisions
                                % 37.10.De Atomcooling methods
\vspace{2pc}
\noindent{\it Keywords}: ultracold collisions, sympathetic cooling, atom trapping, ytterbium
\maketitle

\section{Introduction}
The realization of mixtures of ultracold atomic gases in
conservative atom traps has opened the way for the study of
new fundamental aspects in dilute quantum degenerate systems.
Various combinations of alkali atoms have successfully been
trapped in conservative traps 
and recently even simultaneous trapping of three different alkali
species has been reported \cite{Taglieber2008}. Among the 
experimental achievements in these systems are the realization of
two-species Bose-Einstein condensates \cite{Modugno2002}, quantum
degenerate Bose-Fermi \cite{Hadzibabic2002,Roati2002,Silber2005} and Fermi-Fermi \cite{Taglieber2008,Spiegelhalder2010} mixtures 
and the realization of interspecies Fesh\-bach resonances \cite{Stan2004,Inouye2004,Thalhammer2008, Wille2008,Deh2008,Pilch2009}.
One of the most intriguing aspects of the physics of ultracold mixtures is the preparation of a trapped sample
of ultracold heteronuclear molecules in the rovibrational ground
state, which offers fascinating possibilities to study
ultracold dipolar gases \cite{Baranov2002}, explore new systems
for quantum information \cite{DeMille2002,Micheli2006} or test fundamental
physics \cite{Kozlov1995,Hudson2002}. In bialkali systems several groups have observed ultracold molecules  in the rovibrational ground state in the last few years \cite{Sage2005,Deiglmayr2008, Lang2008,Ni2008} and recently, these studies led to observations of quantum stated controlled chemical reactions \cite{Ospelkaus2010,Knoop2010}.

So far, experimental studies of ultracold mixtures in conservative traps were limited to bialkali  systems. Only recently,
several groups have begun to investigate mixtures of alkali atoms with ytterbium (Yb) and first experimental results of 
magneto-optical trapping \cite{Nemitz2009,Okano2010} and photoassociation \cite{Nemitz2009} have been obtained. Conservative trapping of such mixtures is complicated by the fact that the ground state of Yb is diamagnetic while the ground state of alkali atoms is paramagnetic. Hence, trapping of both species in a common magnetic trapping potential is not possible, in contrast to alkali mixtures where magnetic traps are typically used for evaporative cooling of the mixture. The alternative approach of optical trapping of the mixture is also not straightforward as the polarizability of Yb (and alkaline earth atoms) is significantly smaller than that of alkali atoms for most wavelengths that are typically used for optical trapping. The resulting different trap depths are a severe limitation for many experimental studies.   

In this work, we report on an experimental solution for simultaneous trapping and cooling of alkali and diamagnetic atoms such as Yb in a hybrid trap. 
In our experimental setup, rubidium (Rb) and Yb are held in two separate but spatially overlapping traps. The trapping
scheme uses a clover leaf magnetic trap (MT) \cite{Mewes1996a} for Rb and a bichromatic optical dipole trap
for Yb and thus offers the flexibility to manipulate the two
species independently. As a consequence, the
interspecies interactions can be turned on or off by controlling the spatial overlap of
the two atomic clouds. 

\begin{figure}
\begin{center}
\includegraphics{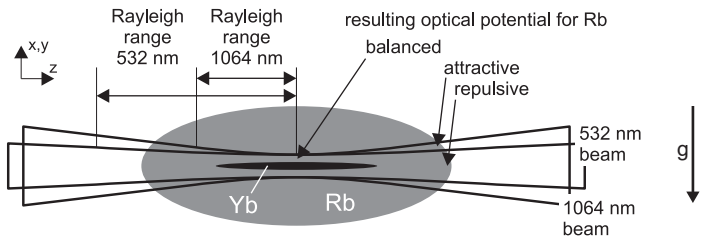}
\caption{\label{fig:combtrap} Scheme of the combined trap with the
spatial configuration of the atomic clouds and the BIODT laser
beams (not to scale).}
\end{center}
\end{figure}

The BIODT for Yb is created by two superimposed laser beams at $532\units{nm}$ and $1064\units{nm}$,
both red-detuned with respect to the dominant  $^1S_0$ $\rightarrow$
$^1P_1$ transition  in Yb at 399\,nm (see Fig.\,\ref{fig:combtrap}). The diamagnetic Yb atoms are
intrinsically not affected by the MT, while for Rb the
optical potentials cancel to lowest order for equal beam waists and an appropriate power
ratio, since the $1064\units{nm}$ light is red-detuned and the
$532\units{nm}$ light is  blue-detuned from the dominant atomic
transition in Rb at 780\,nm.  In this hybrid trap, we have investigated sympathetic cooling of
the five isotopes $^{170}$Yb, $^{171}$Yb, $^{172}$Yb, $^{174}$Yb
and $^{176}$Yb through collisions with $^{87}$Rb in the $(F=1,
m_F=-1)$ state at temperatures of several 10 $\mu$K. A strong
dependence of the interspecies collisional cross section on the
mass of the ytterbium isotope was observed.

A related trapping scheme which uses only a bichromatic
light field for the independent adjustment of the trapping
strengths for two species was proposed by Onofrio et al. \cite{Onofrio2002}. Experimentally, a hybrid trap involving a magnetic trapping potential and a single-color light field has recently been used to study the entropy exchange in an ultracold mixture \cite{Catani2009}.

\section{Experimental setup}

\begin{figure}
\begin{center}
\includegraphics{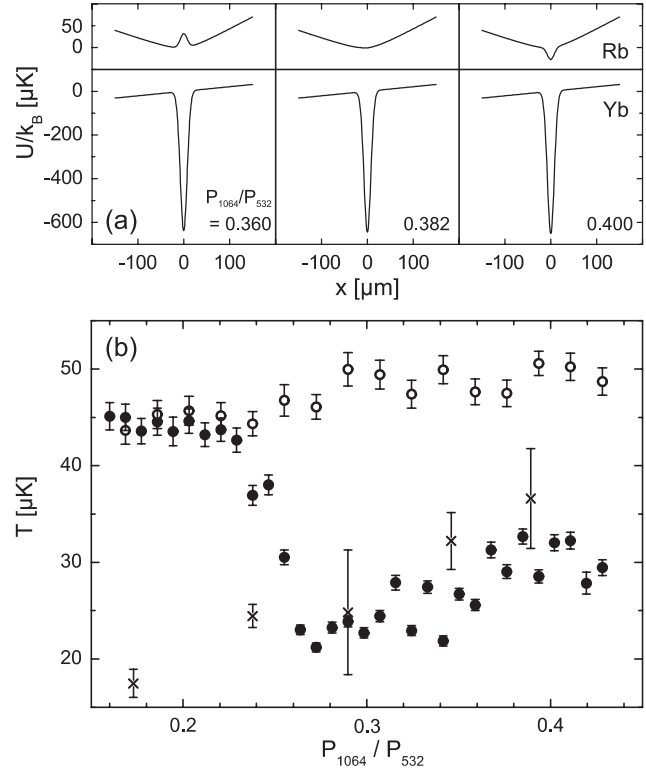}
\caption{\label{fig:powerratio} (a) Calculated potentials of the hybrid trap for
$^{176}$Yb and $^{87}$Rb (in the ($F=1, m_F=-1$) state) using the
experimental parameters in (b) and assuming perfect gaussian laser
beam profiles. The subfigures show radial cuts through the center of the
combined trap including gravity as a function of the ratio $P_{1064}/P_{532}$ (at $P_{532}=3.4 W$). \\(b) Steady state
temperatures (after $1\units{s}$ contact time) in the hybrid trap
as a function of the power ratio of the two light fields. Shown
are the temperatures of $^{176}$Yb with ($\bullet$) and without
($\circ$) Rb present and the Rb temperatures ($\times$). }
\end{center}
\end{figure}

A typical experimental sequence starts with an Yb magneto-optical trap (MOT) on the broad  $^1S_0$ $\rightarrow$
$^1P_1$ transition at $399\units{nm}$ which is loaded from a Zeeman slower. Up to $2\times 10^7$ Yb atoms
 are then transferred into a second MOT
 operating on the narrow $^1S_0$ $\rightarrow$
$^3P_1$ intercombination transition at $556\units{nm}$ where a density in the range of $10^{11}$\,cm$^{-3}$ and a
temperature of $\approx70\units{\mu K}$ are reached. Subsequently, the atoms
are loaded into the BIODT by simply turning off the MOT beams. The
two light fields of the BIODT have 1/e$^2$-radii  of $w_{1064}
\approx w_{532} \approx 15\units{\mu m}$ with a residual
ellipticity  on the order of 10\% and typical laser powers of
$P_{1064}=1.0\units{W}$ and $P_{532}=3.4\units{W}$. Stable overlap
of the two light fields is achieved by an active stabilization of
the relative position. In this configuration, the calculated trap
depth for Yb is $U/k\idx{B} \approx -600\units{\mu K}$ and the
measured trap frequencies are $\omega\idx{r}\approx2\pi\times
3.3\units{kHz}$ radially and $\omega\idx{z}\approx2\pi\times
29\units{Hz}$ axially.  After a time of $\approx20\units{s}$ (which is required for the preparation of Rb), the Yb cloud has reached its equilibrium temperature
of $\approx50\units{\mu K}$ through plain
evaporation. At this stage, up to $2\times 10^5$
atoms at a density of about
$10^{13}\units{cm}^{-3}$ are trapped in the BIODT in the case of $^{176}$Yb. The lifetime in the BIODT of about
$130\units{s}$ is limited by background gas collisions. For
different Yb isotopes, the initial atom number in the optical trap varies
by a factor of 2-3 for technical reasons (mainly determined by the specific natural abundance) while the temperature exhibits no significant isotope dependence.

After loading of the BIODT with Yb, a cloud of ultracold Rb is prepared in the same vacuum chamber. 
In a standard MOT, $10^9$ $^{87}$Rb  atoms at a temperature of several 100\,$\mu$K are loaded from a Zeeman slower. Care is taken to avoid overlap between 
the Rb MOT and the optically trapped Yb atoms as this leads to a significant loss of Yb atoms. After loading of the MOT, the Rb atoms are transferred to the clover leaf magnetic trap with measured trap frequencies
of $\omega\idx{r} = 2\pi\times 175\units{Hz}$ and
$\omega\idx{z} = 2\pi\times 13.5\units{Hz}$. In the magnetic trap, we prepare an ensemble of typically
$10^7$ Rb atoms in the ($F=1, m_F=-1$) state at a temperature of
$\approx15\units{\mu K}$ by radio-frequency induced evaporation. The center of the
MT is initially located at a distance of $0.7\units{mm}$ from the
BIODT. Due to the diamagnetic ground state of Yb, any magnetic
field needed for cooling and trapping of Rb does not interfere
with the optically trapped Yb atoms.  Subsequently, we move the MT
for Rb to the position of the BIODT for Yb within $100\units{ms}$
by applying additional magnetic offset fields. Thus, we can control the beginning and the duration of the contact between the two species. The MT and the BIODT, both have their weak axis
aligned along the $z$-direction to ensure maximum overlap of the
two atomic clouds. All atom numbers and temperatures are determined by standard absorption imaging and time-of-flight techniques.

Figure \ref{fig:powerratio}(a) shows radial cuts through the trap
center of the  calculated total trapping potentials for Yb and Rb
for various ratios $P_{1064}/P_{532}$. The calculation includes
all magnetic and optical potentials as well as gravity which is perpendicular to the axis of the BIODT and the magnetic trap. Though Fig.~\ref{fig:powerratio}(a) is certainly a simplified representation as the trap geometry is reduced to 1-D, it is well-suited for a qualitative discussion of the influence of the power ratio on the hybrid trap. For
$P_{1064}/P_{532}=0.382$, the combined optical potential for Rb
vanishes and the radial confinement is determined by the magnetic
trapping potential alone. Thus, if optical and magnetic potentials
are correctly adjusted, the Yb cloud is trapped in the BIODT at
the same spatial position where the Rb cloud is held magnetically.
For smaller power ratios, the contribution of the $532\units{nm}$ laser 
 to the BIODT potential becomes dominant creating a
repulsive barrier for Rb in the trap center, and the contact between the two species decreases with increasing barrier height. If the power ratio is too high, an
attractive dimple forms for Rb caused by the dominant light shift
of the infrared beam at $1064\units{nm}$. In this case, contact
between Yb and Rb is possible, but the Rb trap is no longer
independent from the Yb trap, and forced  radio frequency evaporation of Rb is
stalled since all magnetic substates are trapped in the attractive
optical potential. 

\section{Results}

In the experiment, we observe thermalization of Yb with the colder
$^{87}$Rb after the two clouds are brought into contact. The time
until a steady state temperature is reached ranges from
$100\units{ms}$ to several seconds and depends on the Yb isotope
and on the number of $^{87}$Rb atoms. In agreement with the discussion
above, we found the power ratio of the two BIODT beams to be of
crucial importance to reach the lowest Yb temperature as can be
seen in Fig. \ref{fig:powerratio}(b) where the temperature of an $^{176}$Yb cloud is plotted as a function of the power ratio for a
fixed contact time. For $P_{1064}/P_{532}$ below
a critical value of $\approx0.25$, the Yb temperature is
unaffected by the presence of Rb, due to the lack of thermal
contact while above this value, Yb thermalizes with Rb. With increasing
power ratio, however, the Rb temperature increases due to the
influence of the resulting attractive BIODT-potential, leading
also to an increase of the steady-state Yb temperature. 

The
coldest mixtures are obtained at $P_{1064}/P_{532} \approx
1.0\units{W}/3.4\units{W} \approx 0.29$. The difference to the value of $0.382$, which results in light shift compensation for Rb in the model potential, may be attributed to uncontrolled deviations of the experimental optical  potential from the calculated one. In addition, cooling of Yb should already be possible for a slightly repulsive potential for Rb if the interspecies collisional cross section is large. To account for this effect in a quantitative way, will require an improved experimental control over the details of the trap geometry. In this context, it is worth noting that the Rb temperature
starts to increase already at a power ratio 
below the critical value. We attribute this to the fact, that the
two light fields are not perfectly overlapping and attractive
regions in the light field are forming even for low infrared
power. This is partly due to a deviation of the beam shapes from a
perfect gaussian shape, but even for perfect gaussian beams, complete
overlap of the two light fields is not possible due to the
different divergences of the two laser beams (see
Fig.~\ref{fig:combtrap}).

\begin{figure}
\begin{center}
\includegraphics{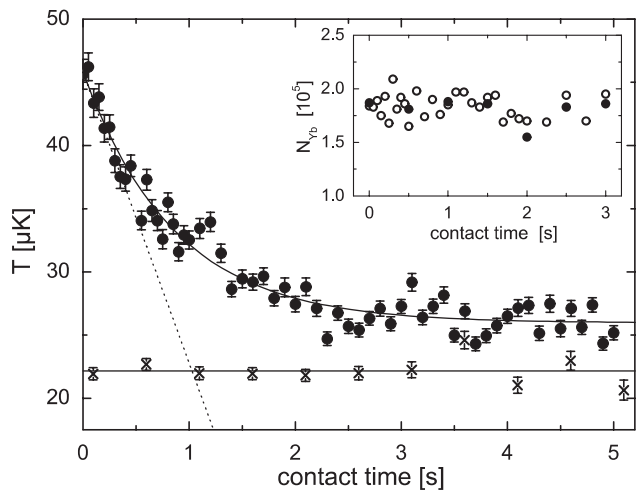}
\caption{\label{fig:thermcurve} Temperature evolution of
$^{176}$Yb ($\bullet$) and $^{87}$Rb ($\times$) during
thermalization. The solid lines are exponential (Yb) and constant
(Rb) fits to the data. The dotted line shows the slope of the
exponential fit at $t=0$ corresponding to the initial cooling rate
of Yb. The inset shows the atom numbers for sympathetically cooled
Yb ($\circ$) and for Yb without Rb present ($\bullet$).}
\end{center}
\end{figure}

We have studied the
thermalization process in the Rb-Yb mixture for the bosonic isotopes $^{170}$Yb, $^{172}$Yb,
$^{174}$Yb and $^{176}$Yb, which posses all a nuclear spin of $I = 0$, as well as for the fermionic isotope
$^{171}$Yb with a nuclear spin of $I = 1/2$.
After the two ensembles are brought into contact, we record the
evolution of the Yb temperature as a function of the contact time.
In Fig.~\ref{fig:thermcurve}, a sample curve for $^{176}$Yb is
shown, which clearly demonstrates sympathetic cooling. We do not observe any
loss of Yb atoms during thermalization, as is shown in the inset of
Fig.~\ref{fig:thermcurve}. In the case of $^{176}$Yb and
$^{174}$Yb, we measure steady state temperatures close to
the Rb temperature, while for the other isotopes the steady-state
temperature stays well above the Rb temperature, which may be
accounted for by a smaller scattering cross section in combination
with additional heating and imperfect spatial overlap of the two
atomic clouds. In the case of $^{170}$Yb we observe almost no temperature 
reduction. 

During the contact time, the Rb temperature is kept constant by
continuously applying a radio frequency of 1.5\,MHz to the
magnetically trapped Rb atoms. Under this condition, we observe a
significant loss of Rb on a timescale of several seconds in the
presence of the BIODT light fields. We attribute this loss to an
imperfect cancellation of light shifts for Rb which results in an
adiabatic compression and heating of the Rb ensemble when the MT
is shifted onto the BIODT. In addition, the Rb atoms are heated by photon scattering 
from the relatively intense BIODT beams. In the presence of a constant
radio frequency both heating mechanisms translate into a loss of atoms. This
interpretation is supported by another series of experiments,
where the radio frequency was turned off during the contact time.
In that case, we found the Rb atom number to stay constant
but observed a rapid initial increase of the Rb temperature to
$\approx 30\units{\mu K}$ within about $130\units{ms}$ followed by
a constant heating of about $1\units{\mu K/s}$ (the latter in
agreement with the expected heating rate due to photon scattering
in the BIODT). Results of both experimental methods have entered
into the quantitative analysis below.

From measurements of the thermalization rate at 
ultracold temperatures as presented in Fig.~\ref{fig:thermcurve}, information about the interspecies
cross section $\sigma\idx{YbRb}$ can be obtained. For a constant Rb temperature and 
$N\idx{Rb}\gg N\idx{Yb}$, the evolution of the Yb temperature can be written as \cite{Mosk2001}
\begin{equation}
\dot{T}\idx{Yb}(t)\, =\, -\gamma\idx{th}\cdot\Delta T (t) +
\dot{T}\idx{heat}\, ,
\end{equation}
with the thermalization rate $\gamma\idx{th}$, 
and $\Delta T (t) = T\idx{Yb} (t) - T\idx{Rb} (t)$ the temperature difference between Yb and Rb. The heating rate for Yb,
$\dot{T}\idx{heat}$, which is taken to be independent of the presence of Rb atoms is
measured to be $0.9\units{\mu K/s}$ and can be accounted
for by photon scattering in the BIODT. Experimentally, we determine the initial cooling rate $\dot{T}\idx{Yb}(0)$ and temperature difference $\Delta T (0)$ and thus the initial thermalization rate $\gamma\idx{th,0} = (\dot{T}\idx{heat} - \dot{T}\idx{Yb}(0))/\Delta T (0)$ can be expressed in terms of experimentally determined quantities. 

Assuming as a first approach a temperature-independent scattering cross section $\sigma\idx{YbRb}$, the
thermalization rate is on the other hand given by 
\begin{equation}
\gamma\idx{th}\,=\,\frac{\xi}{\alpha}\, \gamma\idx{coll}\,=\, \frac{\xi}{\alpha}\, \sigma\idx{YbRb}\,\bar{v}\,n\idx{YbRb}\ .
\end{equation}
Here $\alpha$ is the number of collisions which is required for thermalization if the collisional partners have equal mass \cite{monroe1993,wu1996,arndt1997}, $\xi$ is a mass-dependent reduction factor \cite{Mosk2001}, $\bar{v}=\sqrt{\left(8k\idx{B}/\pi\right)\left(T\idx{Yb}/m\idx{Yb}+T\idx{Rb}/m\idx{Rb}\right)}$
the thermal velocity and 
\begin{equation}\label{equ:overlapdensity}
n\idx{YbRb}\,=\, (N^{-1}\idx{Yb}+N^{-1}\idx{Rb})\int n\idx{Yb}({\bf
r})n\idx{Rb}({\bf r}) {\rm d}^3{\bf r}\,\approx N_{Rb}\,f_{YbRb}
\end{equation}
the overlap density, where we have introduced an overlap function $f_{YbRb}$. Introducing the parameter $\beta = (\dot{T}\idx{heat} - \dot{T}\idx{Yb}(0))\left(\bar{v}_0 N\idx{Rb}\Delta T(0)\right)^{-1}$ with the initial thermal velocity $\bar{v}_0$, the scattering cross section can be written as
\begin{equation}
\sigma\idx{YbRb}\,=\,\frac{\alpha}{\xi}\, \beta\, f_{YbRb}^{-1}\ . 
\end{equation} 

While $\beta$ is completely determined by experimentally measured quantities and $\alpha$ and $\xi$ can be modeled, the overlap function $f_{YbRb}$ is experimentally difficult to determine precisely as it is strongly influenced by small deviations of the trapping potential from the ideal situation. Due to the composition of the trapping potential of two optical fields and a magnetic field in our experiment,  such uncontrolled deviations are experimentally unavoidable, as already mentioned. Thus, precise absolute values of scattering cross sections cannot be determined from our measurements directly. However, it is possible to determine relative cross sections for different Yb isotopes if $f_{YbRb}$ is assumed to be constant. Hence, we define   $\beta\idx{norm}(^{x}$Yb$^{87}$Rb$) = \beta(^{x}$Yb$^{87}$Rb$)/(5.36 \times 10^{-6} \units{m^{-1}})$ corresponding to normalizing the interspecies scattering cross section for the isotope $^{x}$Yb to the experimentally determined value for $^{176}$Yb. 

\begin{figure}
\begin{center}
\includegraphics{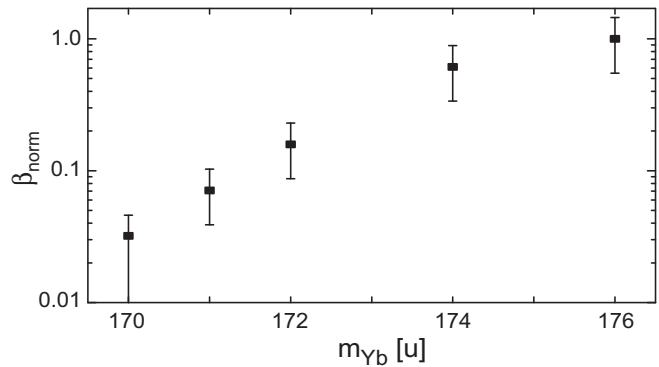}
\caption{\label{fig:results} Coefficient $\beta\idx{norm}$ which is a measure for the relative interspecies scattering cross section between $^{87}$Rb and a given Yb isotope at the experimental temperature of $T\approx 50\mu$K.}
\end{center}
\end{figure}

In Fig.~\ref{fig:results}, the experimentally determined values of $\beta\idx{norm}$ are shown
for all investigated Yb isotopes. The error bars are inferred from the experimental variations of 12 independent measurements of $\beta(^{176}$Yb$^{87}$Rb$)$. These variations are attributed to uncontrolled temporal variations of details of the trap geometry and thus of $f_{YbRb}$. Within experimental uncertainties, $\beta(^{170}$Yb$^{87}$Rb$)$ is compatible with zero. The observed strong mass
dependence of $\beta\idx{norm}$ is consistent with the scattering properties being determined by details of the molecular potential of ground state YbRb, which in turn are strongly influenced by the mass of the atoms \cite{Ferrari2002}.  In recent experimental studies at lower temperatures in the $\mu$K regime, that will be reported in detail elsewhere, we have found that for $^{174}$Yb the scattering cross section increases significantly for decreasing temperatures and becomes much larger than the one for $^{176}$Yb. While this is an obvious violation of the above assumption of a temperature-independent scattering cross section, the values for $\beta\idx{norm}$ in Fig.~\ref{fig:results} are still a good representation of the relative scattering cross sections at the experimental temperature of around 50\,$\mu$K as they have been determined using only the initial slope of the thermalization curve (Fig.~\ref{fig:thermcurve}). However, it should be noted that the height of the centrifugal barrier for p-wave-collisions \cite{Dalibard1999}, which we have calculated using the $C_6$-coefficients for Yb and Rb \cite{Kitagawa2008,Kempen2002,Derevianko2001}, is only around 60\,$\mu$K. Thus, a contribution of p-wave collisions to the collisional properties of the Yb-Rb mixture described here may not be excluded.

In order to give at least an estimate of the magnitude of the interspecies cross section in mixtures of Yb and Rb, we have evaluated $\sigma\idx{^{176}Yb^{87}Rb}$ under the assumption of a perfect trap geometry. Under this condition, we calculate a value for the overlap function of $f_{YbRb} \approx 5\times10^{10}\units{m^{-3}}$ using the experimentally determined temperature and trap parameters of the individual traps. Taking $\alpha\,\approx \,2.7$ as the number of collisions required for thermalization \cite{monroe1993, wu1996,arndt1997} and calculating a reduction factor of $\xi\,=\,0.89$ for the Yb-Rb mixture \cite{Mosk2001} we obtain $\sigma\idx{^{176}Yb^{87}Rb}\,=\,(3.3\pm 1.5) \times 10^{-16}\,\units{m^2}$. Here the error is only statistical and does not include the systematic error due to the overall uncertainties concerning details of the trap geometry which can easily change $f_{YbRb}$ by a factor of 2. Nevertheless, we may still conclude that at a temperature of around 50\,$\mu$K $\sigma\idx{^{176}Yb^{87}Rb}$ is of a similar magnitude as the s-wave scattering cross section for collisions between $^{87}$Rb atoms \cite{Kempen2002}. 

\section{Conclusion}

In conclusion, we have experimentally realized an ultracold
mixture of a paramagnetic and a diamagnetic atomic species and
thus demonstrated the experimental feasibility to investigate ultracold, conservatively 
trapped mixtures which combine atomic species that behave very differently
in external fields. The trapping geometry in which we have
achieved sympathetic cooling of Yb through collisions with Rb may
easily be adopted for other combinations of atomic species with
similar magnetic properties. It allows for nearly independent manipulation of the trapped species and 
thus offers a lot of flexibility for experimental studies. In addition, we have determined 
 elastic scattering properties between $^{87}$Rb and five different Yb
isotopes at $T \approx 50\mu$K and observed a strong dependence on the mass of the Yb
isotope. At least for the isotopes $^{176}$Yb  and $^{174}$Yb, the scattering cross sections are large enough for efficient sympathetic cooling with $^{87}$Rb as a coolant. 

These first results on combined trapping of Rb and Yb
open the prospect to realize a novel kind of quantum degenerate
mixture. Together with recent experimental \cite{Nemitz2009} and theoretical  work \cite{Sorensen2009,Meyer2009} on YbRb molecules, the findings presented in this paper will also be valuable for the development of a strategy to create paramagnetic YbRb ground state molecules. 

We acknowledge financial support from the Deutsche
Forschungsgemeinschaft under the Emmy-Noether program and under
SPP 1116. F.B. was supported by a fellowship from the Stiftung der Deutschen Wirtschaft.\\

%\bibliography{C:/1-Users/axel_work/References/References}
%\bibliographystyle{iopart-num}

\providecommand{\newblock}{}

\end{document}